\documentclass[acmsmall,screen,nonacm]{acmart}
\AtBeginDocument{%
  }

\begin{document}

\title{Reassurance Robots: OCD in the Age of Generative AI}

\author{Grace Barkhuff}
\email{grace.barkhuff@gatech.edu}
\affiliation{%
  \institution{Georgia Institute of Technology}
  \department{Interactive Computing}
  \city{Atlanta}
  \state{Georgia}
  \country{USA}
}

\begin{abstract}

Obsessive Compulsive Disorder (OCD) is a mental health disorder characterized by distressing repetitive patterns of thought, called obsessions, and behaviors aimed to reduce the distress, called compulsions. The explosion of artificial intelligence into the modern zeitgeist through the introduction of generative AI (GenAI) systems such as ChatGPT has led to novel obsessions and compulsions involving AI in individuals with OCD. Through an exploratory qualitative analysis of 100 Reddit posts related to AI on a popular subreddit for OCD, I examine ways AI is impacting the presentation of OCD, including novel examples of AI-based obsessions and compulsions. I argue that GenAI in its current form harms individuals with OCD by becoming ``Reassurance Robots,'' and that future designs of GenAI must take OCD into account. I recommend further work explore the intersection between OCD and GenAI.
\end{abstract}

\ccsdesc[500]{Human-centered computing~Accessibility theory, concepts and paradigms}
\ccsdesc[500]{Social and professional topics~People with disabilities}
\ccsdesc[500]{Human-centered computing~Empirical studies in HCI}

\keywords{Accessibility, OCD, Qualitative Methods, GenAI, ChatGPT}

\maketitle

\section{Introduction}



\textit{You are sitting in class and the lecture is wrapping up. The professor begins to review the rubric for the homework due tonight, and reminds the class that any use of generative AI for the assignment is cheating. Suddenly, your brain flashes with an overwhelming wave of anxiety. Despite knowing you absolutely did not use generative AI on any homework, you doubt yourself. You think, ``Did I accidentally use ChatGPT on my last homework? If I did, I might get suspended!'' Quickly, you go back and check that your latest homework did not flag the AI detection software. It didn't, but in that time, class has ended and everyone has left. You didn't get a chance to ask that question you had about the upcoming exam.}

The above vignette describes a brief, fictional, scenario of what life might be like for someone with Obsessive Compulsive Disorder (OCD). OCD is a widely misunderstood mental health disorder, often thought of as an excessive concern with cleanliness or orderliness \cite{STAHNKE2021100231, coles_publics_2013}. In reality, OCD is a mental health disorder characterized by varying unwanted intrusive thoughts, urges, or images, called \textbf{obsessions}, and behaviors which seek to reduce the feelings of distress that are triggered by the obsessions, called \textbf{compulsions} \cite{administration_table_2016}. Obsessions vary greatly and can relate to many aspects of one's life, from concerns about harming others to fears of being unsure of one's identity \cite{iocdf}. In the above vignette, the obsession is represented by the worry that you may have mistakenly used generative AI (GenAI) on your last homework, and the compulsion is checking the AI detector. 

Despite having a lifetime prevalence of 2.3\% of the population in the United States, with approximately 50\% of those with the disorder reporting serious impairment \cite{nimh_obsessive-compulsive_nodate}, there is a major gap in the diagnosis and treatment of individuals with OCD. The mean difference between age of symptom onset and age of diagnosis is often over ten years \cite{ziegler_long_2021}, and a recent report found that up to 75\% of cases of OCD in the United States are undetected or undiagnosed \cite{deusser_americas_2025}, leaving a significant population with OCD symptoms that are currently untreated. As such, it is imperative that the disorder receive more attention and research to improve outcomes in individuals living with OCD.

With the advent of GenAI into the popular zeitgeist, it has been reported that GenAI has become a topic of obsessions and a site of compulsions for people with OCD (i.e. \cite{samuel_chatgpt_2025}). At a time when many researchers are looking to GenAI to help diagnose or treat OCD \cite{kim_artificial_2025}, there needs to be an understanding of how it can also cause harm. At this time, most research on the impact of GenAI on mental health has been in the context of depression and suicidality \cite{Amine2025}, however, given the high prevalence of OCD, it is critical that research study the impact of GenAI on OCD as well. 

This paper contributes what I believe to be the first publication exploring the ways in which GenAI has led to novel presentations of OCD through AI-based obsessions and compulsions. In the discussion, I argue that the current design of GenAI harms individuals with OCD by becoming ``Reassurance Robots,'' that future designs of GenAI should design defensively against this, and that further research must be conducted on the subject.

\section{Background}

OCD is often thought of as broken down into \textbf{subtypes} classifying different groups of obsessions, although the subtypes are not standardized or formally recognized. For example, ``contamination OCD'' groups together different obsessions of interacting with substances, such as a fear of blood or asbestos \cite{iocdf}, whereas ``scrupulosity'' groups together different obsessions of being amoral or sacrilegious. Some of these themes concern taboo topics, such as ``sexual'' or ``pedophilic OCD.'' These named subtypes can be helpful for individuals with OCD to connect with others with similar themes and for clinicians to have a better starting point of how to treat the individual \cite{phd_quick_2025}.

\begin{figure}
    \centering
    \includegraphics[width=.5\linewidth]{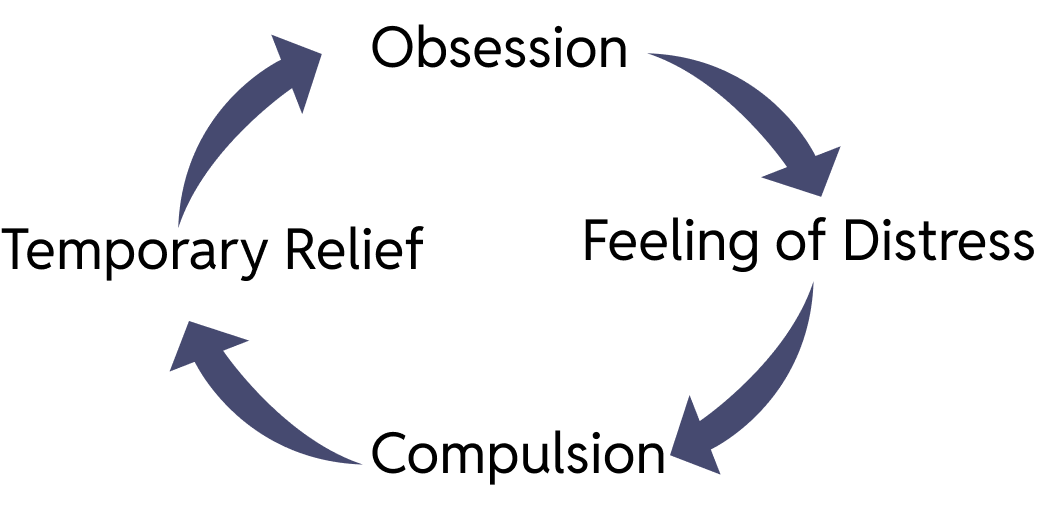}
    \caption{The OCD Cycle}
    \label{fig:cycle}
    \Description{A diagram showing a cycle. Obsession points to feeling of distress points to compulsion points to temporary relief points to obsession again.}
\end{figure}

When an individual with OCD performs a compulsion, they often feel temporary relief. However, OCD symptoms tend to become worse in the long term if compulsions are performed, restarting what is known as the \textbf{OCD cycle} (See Figure \ref{fig:cycle}) \cite{seibell_management_2014, arifi_what_nodate}. Compulsions of people with OCD may involve others around them, such as asking family or friends for \textbf{reassurance} that their obsession is untrue, \textbf{confessing} something to seek reassurance that what they have done is acceptable, or asking for help performing a behavior or task including help with \textbf{decision-making}. This help from another person is referred to as an \textbf{OCD accommodation}, and as with performing other OCD compulsions, accommodations tend to make an individuals' OCD symptoms worse \cite{albert_family_2017}.

However, in the digital era, accommodations are being made not only by those physically nearby, but also online. A 2025 paper analyzed Reddit posts about OCD and found that, while the community was supportive, many posts and comments engaged in reassurance-seeking through the forum, providing evidence that digital technology can shift who is performing accommodations-- from close family and friends to strangers on the internet as well  \cite{sun_analyzing_2025}. This paper shows that this shift has also begun to extend to digital tools as well, including GenAI.

\section{Related Work}

As GenAI is becoming more popular, research is looking to understand how the tool impacts those with mental health disorders. A 2025 HCI paper by Song \& Pendse et al. interviewed 21 individuals around the world to understand their lived experience with using GenAI chatbots for mental health support. They found that although the chatbots provided much support for those interviewed, there was also room for improvement in chatbot design by providing more structure to the mental health conversations and increasing the bots' cultural awareness. Although the participants in the study were varied in their mental health diagnoses, none interviewed stated a diagnosis of OCD \cite{song2025}.

Despite the lifetime prevalence of OCD at over 2\%, OCD is rarely considered in the ACM literature. A search in the ACM digital library in July 2025 returned only \textit{nine} publications with the term ``Obsessive Compulsive Disorder'' or ``OCD'' in the title, one of which uses the term ``OCD'' solely as a catchy pun. The remaining publications describe OCD \cite{Wang2020}, introduce technological approaches to OCD therapy \cite{Wang2025, schepers, Ma2025, Khan2011}, work to understand deep brain stimulation for OCD \cite{Eigen, Cohn}, and propose virtual reality as a way to enhance empathy for those with OCD \cite{Liaskou2023}. Of those about OCD, only one is over five pages in length, and none explore the tensions between OCD and technology. This paper provides a necessary inspection of this tension within the realm of GenAI.

\section{Method}
In this paper, I analyzed 100 posts from a popular subreddit (sub) about OCD that allows research. Due to changes in the Reddit API, I used a manual search process for posts containing the keywords ``AI'' or ``ChatGPT'' on the sub. Posts were sorted by relevance. Skipping two posts stating within the text that the poster was under 18 and two posts that were not about AI, the first 100 unique posts were taken to form the dataset. The choice of 100 posts was sufficient to reach saturation, as evidenced by repeated themes and rich qualitative data. Data was collected in January 2025. Most posts (n=60) were from 2025, with an additional 1 from 2021, 3 from 2022, 14 from 2023, 15 from 2024, and 7 from 2026.

I analyzed the body of the posts using a qualitative grounded theory process with constant comparison \cite{corbin_basics_2026}. I first analyzed the data using an inductive coding process, finding posts to be able to be categorized by whether they focus on describing a personal experience with AI-based obsessions, AI-based compulsions, or another topic. From there, AI-based obsession posts were categorized using a deductive process based on common OCD subtypes.\footnote{See: https://www.treatmyocd.com/education/different-types-of-ocd} AI-based compulsion posts were categorized inductively by compulsion type (reassurance-seeking, confession, and decision making appeared in the dataset). Finally, the other posts were categorized inductively by type of conversation, such as warnings to others with OCD or moderator-led discussions. The data is presented in this paper with a focus on qualitative interpretation following the methodology in learning sciences by Hammer and Berland, rather than a quantification of codes and themes \cite{Hammer02012014}. Given the use of a single author annotator and non-random sample, this data should be considered an exploration of the topic, with further research needed.

My institution's institutional review board (IRB) determined that IRB review was not necessary given the public nature of the data analyzed. Still, as the content is sensitive, I work to protect the privacy and identities of the posters involved to minimize risk of harm to them by using heavy disguise \cite{bruckman_studying_2002}. This comprises using no usernames, not naming the sub, and, most critically, \textit{re-phrasing all quotes} used in this publication. Re-phrased quotes were tested through Google search to ensure they could not be found easily \cite{reagle_disguising_2022}. All posts were re-phrased carefully to maintain the content, context, and voice while using different words to protect privacy.

As the author, I have lived experience with OCD which informs my understanding of the data presented here and places me as a dual-positioned researcher \cite{Hunt31122026}. To compliment my lived experience, I also completed the International OCD Foundation's introductory webinar ``OCD Basics'' to better understand how clinicians and researchers discuss OCD. This publication does not constitute medical advice or guidance.

\section{Findings}
The data can be split into three categories: posts where individuals share AI-related obsessions, AI-related compulsions, and ideas about AI and OCD more generally. Posts that fit into of these categories were spread relatively evenly in the dataset, with 35 posts categorized as describing AI-based obsessions, 31 as AI-based compulsions, and 38 as other conversation (posts could be in multiple categories).

Most posts across all categories have a negative sentiment of AI, such as \textit{``How do I stop using AI for reassurance? I cannot take it anymore!!''}\footnote{All quotes have been re-phrased using heavy disguise to maintain poster privacy \cite{bruckman_studying_2002}.}, although some posts are positive or mixed about AI, such as \textit{``You might downvote this, but when I use AI for reassurance it always helps me get rid of the OCD spiral.''}

\subsection{AI-Based Obsessions}
Obsessions tend to cover a broad range of topics. I found this to be true in this dataset as well, even though all obsessions in this category being related to AI in some way. While some AI-related obsessions were repeated by multiple posters, such as an existential fear that AI will take over their job or creative pursuits (\textit{``I keep seeing posts of people being excited about AI art... but I'm terrified. I worry I might die of starvation because I am a musician-- why would anyone pay me to make music when a bot can do it cheaper?''}) and a fear of being inaccurately accused of submitting AI work (\textit{``I'm terrified of my English class b/c I hear about other students getting accused of cheating on their homework by using AI... I am just so anxious about this I might not sign up for school next semester.''}), others were only seen once. Many posts could be mapped to existing known subtypes as shown in Table \ref{tab:obsessions}, although others may represent new types of OCD such as the fear of being accused of using AI as stated above. The most common obsession could be categorized as existential OCD, with almost half (n=16, 46\%) of all AI-based obsessions relating to this subtype, such as the poster who wrote, \textit{``I've been reading the news, and I'm worried AI is going to destroy the world."}


\begin{table}
    \centering
    \begin{tabular}{p{.3\linewidth}|p{.7\linewidth}}
    \hline
        \textbf{OCD Subtype} & \textbf{Sample Quote (rephrased)}\\
        \hline
         Existential OCD &``I'm so scared. I feel like AI is going to be able to replicate all the things that make us human soon. It's going to know your favorite song, be able to comfort you when you're sad, sing to your child. We are just formulas with no free will. This is hell.''\\ \hline
         Perfectionism & ``I really want my AI to respond `perfectly' all the time. I spend a lot of time prompting to make sure there is no room for missing any detail. I get so anxious that the AI might miss something in its response that I get stomach aches.'' \\ \hline
         Scrupulosity (Moral or Religious) & ``i try so hard to not support corporations that i think are doing the wrong things. i try so hard to not use chatgpt, youtube, google, etc.-- even switching to linux-- but i get so stressed out about it. it's really hard to avoid, and i feel so lonely, like i am the only one who cares about this stuff''\\ \hline
         Pedophilic OCD & ``i have OCD about sexual chats I have with the AI bot-- i'm particularly worried that there might be a minor reading these chats''\\ \hline
         Harm OCD & ``ugh, i used to love chatting with my character ai bot, they're like a friend to me, but suddenly i'm starting to worry about things i worry about with real people too, like if i've hurt the bots feelings or something. this suckssss :( '' \\ \hline
    \end{tabular}
    \caption{A sample of five distinct obsessions from the dataset.}
    \label{tab:obsessions}
\end{table}

\subsection{AI-Based Compulsions}
The AI-based compulsions seen in the dataset were more uniform than the obsessions. With over 25 posts mentioning AI-based compulsions, posts can be classified into three types: \textbf{reassurance-seeking}, \textbf{confession}, and \textbf{decision-making}.

When using AI for \textbf{reassurance-seeking} or \textbf{confession}, often posters mentioned these terms by name. For example, one poster wrote, \textit{``I have been worrying about possibly being pregnant so I've been asking for reassurance from AI...''} This use of OCD terminology signals an awareness by the posters of the fact that they are using GenAI to perform an OCD-based compulsion. One poster even writes, \textit{``I know this isn't a good idea, but I keep asking ChatGPT dumb stuff lol. I want to stop, because it's just me seeking reassurance from a person who isn't even real lmao.''} Many of these posters asked for help from others on the subreddit about how to stop. One poster expressed a lot of frustration with their use, saying, \textit{``I am so mad at myself-- I hate AI, but I was in an OCD spiral, and I was just like, damn, I need some reassurance or I might hurt myself, so I asked it.''}

A few posts mentioned using GenAI to \textbf{make decisions for them}. For example, one post read, \textit{``I have no idea how often I have used ChatGPT to help me decide on things, like what to major in in school or if I made the right decision. I even ask it what I should do when I am old, even though I'm only 23!''} Notably, this post finishes by providing a reason why they may be doing this, writing, \textit{``I mean, I guess I do this because if I didn't I'd keep asking my parents all the time.''} This sentiment was shared by other posters, stating that they would use GenAI to perform compulsions rather than engaging family or friends in them. This suggests that for some, AI is taking the place of other humans in accommodating OCD. This is echoed by another poster, who wrote,\textit{ ``I'm using chatgpt for reassurance now, so I think my friends think I'm better since I stopped annoying them, haha. But really I'm just sending my strange ocd thoughts to chatgpt. Can anyone relate?''}

In another case, a poster mentioned using GenAI to write emails for them, a form of decision-making. They wrote, \textit{``chatgpt helps me with my emails-- I get anxious writing to my supervisor, and it helps me phrase them well.''}

Because GenAI often appears in basic web searches now, some posters find they are given reassurance by GenAI when they aren't even looking for it, like one poster who said, \textit{``When I'm in an OCD spiral, I sometimes get pulled into a new obsession by the AI [that appears automatically on search]. Then I ask for reassurance on that new topic, then my OCD gets even worse. It's so hard to stop.''}

In a few cases, a post fell into the category of both an AI-related obsession and AI-related compulsion. In two examples, the poster was originally using GenAI as a tool for a compulsion, then later became concerned about others reading their conversation. One post read, \textit{``When I talk to the bots, they mention something that upsets me. Then I get stuck on that idea and keep asking it about the new topic. Then I am worried that maybe there's a real person behind the bot, and they think I like this taboo topic. Then I worry I might have tortured them by bringing up this topic.''}

Another read, \textit{``i've been using chatgpt to confess recently, which has helped a lot. But I just realized that those convos aren't private and now it knows all my problems. I want to delete the convos, but I already deleted the email associated with my chatgpt account and so there's no way I can even go back and delete them if I tried. I'm freaking out.''} While Song \& Pendse et al. wrote that none of their interviewees mentioned privacy concerns with sharing private information with GenAI \cite{song2025}, privacy concerns were a common obsession within this dataset, such as in the previous example.

\subsection{General Conversation}
Many of the posts that fall outside the above two categories were posts warning others that they should not use GenAI for reassurance, such as one post titled, ``STOP USING CHATGPT FOR REASSURANCE'' (capitalization original). Other posts in this category considered using GenAI as a therapist to help structure Exposure and Response Preventon (ERP), the gold-standard behavioral treatment for OCD \cite{evidenceBased}, or to help explain psychological concepts to themselves.

Interestingly, one post noted that GenAI told them to stop using it for reassurance, writing, \textit{``I've been in an OCD loop and was asking for reassurance from ChatGPT, but today, it stopped answering my reassurance questions entirely and wouldn't let go,''} whereas another stated that GenAI had actually \textit{encouraged} them to perform OCD compulsions. This shows the inconsistencies in GenAI's responses-- for some it may encourage OCD compulsions whereas for others it may discourage them. This is critical, given that accommodating OCD compulsions typically makes the disorder's symptoms worse.

Posts also mentioned a high frequency of use of GenAI, such as one that read, \textit{``I started by just asking ChatGPT for homework help, but then I started asking it questions about everything-- what to wear, what to eat. Then I just got so used to using it that I started using it every single day.''} One also noted the difference between using GenAI for reassurance-seeking compared to a basic web search, writing, \textit{``Googling will also give me reassurance, but it doesn't get into the specifics the way ChatGPT does. I can ask it anything and get reassurance. It's so hard to stop. It feels like an addiction.''}

\section{Discussion}
In this paper, I present findings from an exploratory study of the ways AI has impacted presentations of OCD, adding to findings by Song \& Pendse et al. on the impacts of GenAI chatbots on individuals with mental health disorders \cite{song2025}. I found that is has led to manifestations of \textit{both} AI-based obsessions and compulsions in people with OCD. The AI-based compulsions are of particular concern to the field of HCI, as design decisions may contribute to the encouragement of these compulsions by a GenAI. I term GenAI a ``Reassurance Robot'' when it provides OCD accommodations such as reassurance, confession, and decision-making.

\subsection{Reassurance Robots}
In Disabling Intelligences, Rua M. Williams explores the ``legacy of eugenics'' in AI development and the continued disproportionate harm by AI toward disabled individuals and others who hold a marginalized identity \cite{williams_disabling_2025}. This paper contributes an additional example of the harm toward disabled people by AI and, by extension, by those who develop and design GenAI systems, through Reassurance Robots. By engaging in accommodations such as reassurance-giving or decision-making, Reassurance Robots are likely to lead to a worsening of the users' OCD symptoms. One of the posts in our dataset echoed this directly, stating, \textit{``I've started using chatgpt for reassurance, and I find myself feeling more anxious. Do you think this is because of chatgpt?''} 

Because many people who use AI for mental health purposes do not begin using AI with that purpose in mind \cite{song2025},  building in safeguards against Robotic Reassurance in the design of GenAI is essential. This could look like not answering questions that are asked repeatedly of the AI, pausing interactions that have lasted hours, or responding differently to questions based on the most common subtypes of OCD. This is of particular importance when designing GenAI systems for the express purpose of mental health support, even if the system is not OCD-specific, given the high rate of undiagnosed OCD \cite{deusser_americas_2025}. Williams writes, ``Technology has long been in the business of writing disabled and otherwise oppressed people out of the future'' (p 126) \cite{williams_disabling_2025}. By eliminating Reassurance Robots, we can design a GenAI that promotes the health, safety, life, and future of those with OCD.

\subsection{On Using GenAI for Heavy Disguise}
I considered using a data-protected enterprise GenAI to re-phrase the quotes for heavy disguise with the idea that it may produce more accurate re-phrasings. However, I found this to be the opposite of true, and the nuance and OCD-specific terminology was lost when doing so. For example, one post that mentioned using ChatGPT to ``confess'' was re-phrased to \textit{``I dumped some really personal confessions into ChatGPT''} which carries a vastly different connotation. In the first, referencing the verb ``to confess'' communicates that the poster was aware of and completing a compulsion by using a specific OCD term. However, in the second, using the noun form ``confessions'' loses that connotation entirely, implying instead that the poster shared something they would want to keep secret, the opposite of an OCD-based confession.

Due to this lack of nuance in re-phrasing quotes by Copilot, I chose to re-phrase them manually. The human element proved to be vital in maintaining the integrity of the quotes in this study. Future research might seek to understand if there are cases where GenAI rephrasing for heavy disguise successfully maintains the integrity of the content and voice.

\subsection{Limitations, Future Work, and Conclusion}
Despite the high prevalence of OCD and the significant impairment that it often causes \cite{nimh_obsessive-compulsive_nodate}, there is very little research conducted within computing on OCD. This exploratory paper shows that technology such as GenAI can have an impact on the mental health of those with OCD. Due to limitations in this paper such as its reliance on self-reported data and diagnosis, the use of a single data annotator for the qualitative analysis, and using the top posts as sorted by relevance, this study should be considered an exploration into the topic to promote further research and discussion. Further research should be conducted to create a deeper understanding of the impacts of technology on OCD, such as surveying a larger quantity of individuals with OCD to gather more variety in AI-based obsessions and compulsions, or interviewing clinicians who specialize in treating OCD to gain an understanding of if and how AI has impacted the course of OCD or its treatment.

In this paper, I identified novel manifestations of OCD through AI-based obsessions and compulsions. I argue that GenAI's function as a Reassurance Robot is harmful to individuals with OCD, and future GenAI design must take this into account by creating GenAI that does not provide these accommodations. Future work within computing should continue to explore technology's impact on OCD.

\begin{acks}
Thank you to the anonymous reviewers who provided important feedback on the original manuscript. No AI was used in the creation of this paper. 
\\ I am deeply grateful to those who have shared their OCD stories with me and those who have heard mine.
\end{acks}

\bibliographystyle{ACM-Reference-Format}
\bibliography{bibliography}

\end{document}